\documentclass[letterpaper, 10pt, conference]{ieeeconf}    

\IEEEoverridecommandlockouts   

\usepackage[utf8]{inputenc}
\usepackage[T1]{fontenc}
\usepackage[scaled=0.95]{inconsolata}

\usepackage{latexsym,amsmath,amssymb,amsfonts,mathrsfs,mathtools}
\usepackage{arydshln}   

\usepackage{amsthm}
\usepackage{empheq}     

\usepackage{algorithm2e, algpseudocode}
\makeatletter
\renewcommand{\ALG@beginalgorithmic}{\small}
\makeatother

\usepackage{float, graphicx, caption, subcaption}
\usepackage[usenames,dvipsnames]{xcolor}
\makeatletter
\let\NAT@parse\undefined
\makeatother
\usepackage[colorlinks=true,linkcolor=magenta,citecolor=blue,
            urlcolor=cyan,filecolor=red]{hyperref}
\usepackage{cite}

\newcommand{\ud}{\mathrm{d}}

\newcommand{\norm}[1]{\left|  #1\right|}

\newcommand{\Ex}{\mathbb{E}}

\newcommand{\bm}[1]{ \begin{bmatrix} #1 \end{bmatrix}}

\theoremstyle{definition}

\theoremstyle{remark}
\newtheorem{remark}{Remark}

\title{\LARGE \bf
  Linear system identification from ensemble snapshot observations
}
\author{Atte Aalto and Jorge Gon\c{c}alves
  \thanks{AA was supported by ERANET for Systems Biology ERASysApp and Fonds National de la Recherche Luxembourg, project CropClock, grant reference INTER/SYSAPP/14/02, and University of Luxembourg Internal research projects PPPD and OptBioSys. JG was partly supported by the 111 Project on Computational Intelligence and Intelligent Control under Grant B18024.}
  \thanks{Both authors are with Luxembourg Centre for Systems Biomedicine;
    University of Luxembourg;
    6 Avenue du Swing; 4367 Belvaux; Luxembourg. (email: atte.aalto@uni.lu and jorge.goncalves@uni.lu)}%
}

\begin{document}

\maketitle
\thispagestyle{empty}
\pagestyle{empty}

\begin{abstract}

Developments in transcriptomics techniques have caused a large demand in tailored computational methods for modelling gene expression dynamics from experimental data.
Recently, so-called single-cell experiments have revolutionised genetic studies. These experiments yield gene expression data in single cell resolution for a large number of cells at a time. However, the cells are destroyed in the measurement process, and so the data consist of snapshots of an ensemble evolving over time, instead of time series. 
The problem studied in this article is how such data can be used in modelling gene regulatory dynamics.
Two different paradigms are studied for linear system identification. The first is based on tracking the evolution of the distribution of cells over time. The second is based on the so-called pseudotime concept, identifying a common trajectory through the state space, along which cells propagate with different rates. Therefore, at any given time, the population contains cells in different stages of the trajectory.
Resulting methods are compared in numerical experiments.


\end{abstract}


\section{Introduction}

Introduction of high--throughput sequencing technologies has caused an increase in produced gene expression data, and even time series data have become widely available. This has raised computational modelling of genetic systems to the pinnacle of today's research in biology. The cost of collecting data is still very high compared to mechanical or electrical systems, for example, and so the gene expression time series tend to be short in length and the sampling frequency low, which has created a demand for tailored methods taking into account the limitations in the data.

Recent years have witnessed another revolution in sequencing technologies. With so-called single-cell techniques, it is possible to obtain gene expression measurements at the level of one cell instead of a population average obtained by traditional batch techniques. Unfortunately, the cell is destroyed in the measurement process, and therefore it is possible to get only one measurement per cell --- albeit from a large number of cells at a time. The amount of data is orders of magnitude larger than with batch experiments, but the obtained ensemble snapshot data call for new modelling approaches. In this paper, we consider this problem from the point of view of linear system identification. Although simplistic, the goal of this work is to obtain evidence on the suitability of the overall strategies for tackling the problem.

A typical single-cell experiment is carried out as follows. The considered cell population consisting of $N=\sum_{j=0}^m N_j$ cells normally originates from a clonal population so that the cells can be expected to behave similarly. At time $T_0 = 0$, a sub-population of $N_0$ cells is measured. At the same time, remaining cells are perturbed somehow, depending on the experiment, for example by introducing a drug or some other stimulant. At later times $T_j$, sub-populations consisting of $N_j$ cells are measured. In the end, the measurement data consist of $m+1$ snapshot observations of ensembles, $Y=\{Y_0,Y_1,...,Y_m\}$, where $Y_j = \big[y_1^{(j)},...,y_{N_j}^{(j)}\big] \in \mathbb{R}^{n \times N_j}$. The vector $y_k^{(j)} \in \mathbb{R}^n$ consists of gene expression levels of $n$ (interesting) genes in the $k^{\textup{th}}$ cell measured at time $T_j$. For a review on single-cell experimental techniques and a discussion on their potential, we refer to \cite{revolution}.

In this paper, we consider linear system identification from data mimicking a single-cell experiment. Assume that the gene expression dynamics of the cell $k \in \{1,...,N_j\}$ in the sub-population $j \in \{0,...,m\}$ are governed by
\begin{equation} \label{eq:lin_dyn}
\ud x_k^{(j)} = \gamma_k^{(j)} Ax_k^{(j)} \ud t + \ud u_k^{(j)}, \qquad x_k^{(j)}(0) \sim P_0
\end{equation}
where $A$ is a sparse matrix (since the dynamics of one gene are known to be influenced by only few other genes), and $u_k^{(j)}$ is a noise process modelled as Brownian motion. The time-scaling constant $\gamma_k^{(j)}>0$ models the development rate of the cell, which varies from cell to cell. The initial state is a random variable with probability distribution $P_0$. The measurement obtained from this cell is 
\[
y_k^{(j)} = x_k^{(j)}(T_j) + v_k^{(j)},
\]
where $v_k^{(j)}$ is measurement noise, and $T_j$ is the measurement time. The assumption of a full-state measurement is of course a simplification, but it is a rather typical one in genetic applications, made to avoid overfitting.

The problem is to estimate the (sparse) matrix $A$ from the ensemble snapshot data $Y$. We introduce two different paradigms for approaching the problem, and develop one method within each paradigm. Firstly, we develop a method based on tracking the propagation of the (probability) distributions of cells over time, and finding a sparse matrix $A$ that produces such propagation.
The second paradigm is based on the so-called pseudotime concept \cite{monocle,GP_pseudotime,Branch_GP}. The underlying idea in this concept is that cell dynamics are not identical through the population, and in particular, some cells develop faster than others. Therefore, the distribution of measurements at time $T_j$ contains information from different developmental stages. Pseudotime refers to the stage of the cell in the development process. In the example \eqref{eq:lin_dyn}, the pseudotime of the measurement of cell $k$ measured at time $T_j$ roughly corresponds to $\gamma_k^{(j)} T_j$. Pseudotime methods infer the developmental stage of each measured cell. In \cite{Bayesian_var} we developed a method for estimating the zero structure of the dynamics matrix $A$ from time series data. Here this method is modified to include an additional estimator for the pseudotime for each measurement, which is carried out simultaneously with the zero structure inference. In the context of gene expression modelling, the zero structure of $A$ can be interpreted as the gene regulatory network (GRN). GRN inference is one of the cornerstone problems studied in systems biology \cite{Bayesian_var,GPDM_GRN,SC_GRN_review_Aerts,SC_GRN_review_Babtie,SCENIC,Mechanistic_SC_GRN}. The two developed methods are compared to the method presented in \cite{Bayesian_var} applied on the time series data consisting of the averages of sub-populations $Y_j$. This corresponds to a traditional gene expression measurement producing a short time series.

Introduction of single-cell sequencing techniques has lead to emergence of methods analysing the resulting data. Methods that infer cell dynamics from such data include \cite{ensemble,population_density_estimation,Hasenauer}. The first two works are concerned with estimating the state distribution from incomplete measurements. The article \cite{Hasenauer} introduces a method for obtaining distributions of unknown parameters in a chosen dynamical model from the measurement distributions. Optimal mass transport has been applied to single-cell data in \cite{SC_OMT2} for reconstructing cell trajectories. GRN inference from single-cell data has been discussed in \cite{SC_GRN_review_Aerts,SC_GRN_review_Babtie}. Inference is done, for example, using gene expression correlations \cite{SCENIC}, or by considering stationary distributions arising from a mechanistic model \cite{Mechanistic_SC_GRN}.

\section{Methods}

The three methods in the comparison are presented in this section. The first, distribution-based method is completely new, and the second, pseudotime-based method is a modification of our earlier method using time series data \cite{Bayesian_var}. The third method is our original method (without the modification) applied on the population average, which corresponds to data obtained from a classical batch experiment. The distribution-based method is estimating the full matrix $A$, whereas the method in \cite{Bayesian_var} is developed for inferring the GRN, that is, the zero structure of $A$. In Section~\ref{sec:example}, the methods are compared in the GRN inference task.

\subsection{Distribution-based method}

The dynamics equation \eqref{eq:lin_dyn} defines the cell trajectory as a stochastic process (if also the development rate $\gamma_k^{(j)}$ is a random variable). At time $T_j$ the cell state has a certain probability distribution $P_j$ (finite-dimensional distribution of the stochastic process), and the measurements $Y_j$ are regarded as samples drawn from this distribution.
The idea is to find a matrix $A$, such that the pushforward measure $e^{A(T_1-T_0)}P_0$ would be close to $P_1$, and similarly for all $j \in 1,...,m$, the pushforward measure $e^{A(T_j-T_{j-1})}P_{j-1}$ should be close to $P_j$. The ``closeness" is measured by the Jensen--Shannon divergence between the two distributions \cite{JenShan} (see Remark~\ref{rmk:KL}). The Jensen--Shannon divergence is defined through the Kullback--Leibler divergence as
\[
\textup{JS}(p \, |\!| \, q) = \frac12 \textup{KL}(p\, |\!| \,m) + \frac12 \textup{KL}(q\, |\!| \, m)
\]
where $m=\frac12(p+q)$. Recalling the definition of the Kullback--Leibler divergence, the Jensen--Shannon divergence can be expressed as
\begin{align} \nonumber
 & \textup{JS}(p \, |\!| \, q) =  \frac12 \int \log(p(x))p(x)\ud x    \\ \label{eq:JS} & \quad + \frac12 \int \log(q(x))q(x)\ud x  - \int \log(m(x))m(x)\ud x.
\end{align}
As opposed to the Kullback--Leibler divergence, the Jensen--Shannon divergence is symmetric with respect to $p$ and $q$. In addition, there is no absolute continuity requirement between the measures correponding to $p$ and $q$. The continuity requirement for Kullback--Leibler divergence is always satisfied, since $m(x) = 0$ implies $p(x)=0 \textup{ and } q(x)=0$.

The identification task can then be formulated as an optimisation problem
\begin{equation} \label{eq:opt_prob}
\min_A \  C(A) + \sum_{j=1}^m \textup{JS}\left(e^{A(T_j-T_{j-1})}P_{j-1} \, \big|\!\big| \, P_j \right)
\end{equation}
where $C(A)$ is some sparsity promoting regulariser, for example $C(A) = \lambda \sum_{i,j}|A_{i,j}|$ is used in our numerical experiment, corresponding to the well-known Lasso approach~\cite{lasso}. 

The optimisation problem in \eqref{eq:opt_prob} is defined for full distributions $P_j$, but the data consist of samples from those distributions. Therefore integrals of the form $\int \log(p(x))p(x)\ud x$ have to be approximated using samples $x_1,...,x_L$ drawn from $p$. Two different approximations for the distribution $p$ are used. The latter $p(x)$ in the integral is approximated by a sum of Dirac delta distributions at the sample points, transforming the integral into a sum (see Remark~\ref{rmk:quadrature})
\begin{equation} \label{eq:int_app}
\int \log(p(x))p(x)\ud x \approx \frac1{L} \sum_{j=1}^L \log(p(x_j)).
\end{equation}
The remaining $p(x)$ is approximated with a Gaussian mixture
\begin{equation} \label{eq:mixture}
p(x) \approx \frac{1}{L(2\pi q)^{n/2}} \sum_{j=1}^L \exp\left( -\frac{\norm{x-x_j}^2}{2q} \right)
\end{equation}
where $q$ is a design parameter. Inserting this into \eqref{eq:int_app} gives
\begin{align} \nonumber 
 &\int \log(p(x))p(x)\ud x  \\ \label{eq:JS_app} & \approx \frac1{L}\sum_{j=1}^L \log\left( \sum_{k=1 \atop k \ne j}^L  \exp\left(-\frac{\norm{x_k-x_j}^2}{2q} \right) \right)+C
\end{align}
where $C=-\log(L(2\pi q)^{n/2})$ and the $k=j$ term has been excluded from the sum, since otherwise the method seemed to give too little weight to measurements on the outskirts of the distribution.

Practical implementation of the method is sketched in Algorithm~\ref{alg:distr}, where the optimisation problem \eqref{eq:opt_prob} with approximation \eqref{eq:JS_app} is solved using simulated annealing. The approximated Jensen--Shannon divergence $\widetilde{\textup{JS}}(X |\!| Y)$ for $X \in\mathbb{R}^{n \times m_X}$ and $Y \in\mathbb{R}^{n \times m_Y}$ is computed as follows. The first term in \eqref{eq:JS} is computed by inserting $X$ into \eqref{eq:JS_app}, the second term by inserting $Y$, and the last term by inserting $[X,Y]$.

The parameter $q$ in \eqref{eq:mixture} and \eqref{eq:JS_app} was chosen differently when computing different terms of the sum in \eqref{eq:opt_prob}. For the $j^{\textup{th}}$ term in the sum, it was chosen as one tenth of the average of the values $|y_i^{(j)}-y_k^{(j)}|^2$ for $i,k \in \{1,...,N_j\}$ and $i \ne k$.

\begin{algorithm}[h]
 \For{$i=1,...,n_{its}$}{
  Draw $\hat A = A^{(i-1)}+\epsilon_i \cdot \textup{randn}(n,n)$\;
  Set $J=C(\hat A)$\;
\For{j=1,...,m}{
Compute $X_j=e^{\hat A(T_j-T_{j-1})}Y_{j-1}$\;
Set $J=J+\widetilde{\textup{JS}}(X_j |\!| Y_j)$\;
}  
\eIf{$\exp\big((J_{\textup{old}}-J\big)/\textup{Temp}_i)>\textup{rand}$}{
set $J_{\textup{old}}=J$\;
set $A^{(i)}=\hat A$\;}{
set $A^{(i)}=A^{(i-1)}$\;}}
\vspace{1mm}
 \caption{The distribution-based method. The simulated annealing temperature $\textup{Temp}_i$ and step size $\epsilon_i$ decrease as the iterations proceed. Here rand and randn denote random variables from the uniform distribution $U(0,1)$ and the normal distribution $N(0,1)$, respectively.}
 \label{alg:distr}
\end{algorithm}

\begin{remark} \label{rmk:KL}
In our experiments, also $\textup{KL}(p|\!|q)+\textup{KL}(q|\!|p)$ was tried as a distance measure between distributions, but the Jensen--Shannon divergence seemed to produce slightly better results.
\end{remark}

\begin{remark} \label{rmk:quadrature}
The approximation \eqref{eq:int_app} can also be obtained by immediately replacing $p(x)$ by the Gaussian mixture \eqref{eq:mixture}, and then approximating the integral using Gauss--Hermite quadrature with only one sample point per one Gaussian distribution in the mixture sum. A better result could perhaps be obtained by using more quadrature sampling points, but this would slow down the computations somewhat, in particular if the dimension $n$ is big.
\end{remark}

\subsection{Simultaneous estimation of pseudotime and the matrix $A$} 

A method for estimating the zero structure of the matrix $A$ from time series data was developed in \cite{Bayesian_var}. The method is based on Bayesian analysis and MCMC sampling. The method also samples the continuous-time trajectory $x$ underlying the sparsely sampled time series data. Similarly, in the pseudotime concept, it is assumed that the measurements are produced by a continuous trajectory $x$, along which the cells propagate with different rates. In this section, a variant of this method will be developed, where also the pseudotimes related to the measurements are estimated simultaneously. To put briefly, the modification made to the method in \cite{Bayesian_var} is that an additional MCMC sampler is constructed for the pseudotime. In this case, the measurement distribution, given the continuous trajectory $x$, is $y_k^{(j)} \sim N(x(\tau_k^{(j)}),R)$ where $\tau_k^{(j)}$ is the pseudotime corresponding to the measurement $y_k^{(j)}$. In \cite{Bayesian_var} the measurement time was fixed and the measurements readily formed a time series, and the measurement distribution was $y_j \sim N(x(t_j),R)$ where $t_j$ was the measurement time of $y_j$.

In this method, an indicator variable is introduced for the zero structure of the matrix $A$. That is, the element $A_{i,j}$ is represented as a product $A_{i,j}=S_{i,j}H_{i,j}$ where $S_{i,j} \in \{0,1\}$ is an indicator variable indicating whether the element $(i,j)$ is zero or not, and $H_{i,j} \in \mathbb{R}$ is the magnitude variable. The object of interest is then the posterior distribution of the indicator variable $S$, given the data $Y=[Y_0,...,Y_m]$, and the corresponding measurement times $T=\{T_0,...,T_m\}$: 
\begin{align*}
& p(S|Y,T) \propto p(Y|S,T)p(S) \\ & = p(S) \iiint p(Y,x,H,\tau|S,T) \ud x \, \ud H \, \ud\tau  \\ &= p(S) \iiint p(Y|x,\tau)p(x|S,H) p(\tau|T) p(H) \ud x \, \ud H \, \ud\tau
\end{align*}
where we have first used the Bayes' law, then introduced the latent variables $x$, $\tau$, and $H$, where $\tau = \{ \tau_1^{(0)},...,\tau_{N_0}^{(0)},...,\tau_1^{(m)},...,\tau_{N_m}^{(m)}\}$ is the pseudotime vector, $H$ is the magnitude variable, and $x$ is the continuous trajectory. Finally, the probability chain rule is applied to obtain known distributions. The measurement model is 
\[
p(Y|x,\tau)=\prod_{j=0,...,m \atop k=1,...,N_j} N\left(y_k^{(j)} ;  x(\tau_k^{(j)}),R\right), 
\]
that is, it is assumed that the measurements are obtained from the same trajectory at different developmental stages, which is not the same as the true measurement time.

The integral with respect to the magnitude variable $H \in \mathbb{R}^{n \times n}$ is possible to carry out analytically (it is done in \cite{Bayesian_var}), assuming that the rows of $H$ are normally distributed $H_{i} \sim N(0,M_i)$, and independent. A time interval $[\underline{T},\overline{T}]$ is defined for the continuous trajectory. The pseudotimes should be contained in this interval. The integral is
\begin{align*}
&\int p(x|S,H)p(H)\ud H    
  \\ & \propto \prod_{i=1}^n  \frac{\exp(\Phi_i(x))}{\big|M_i[S_i]^{-1}+\frac1{q_i}\mathbb{X}[S_i] \big|^{1/2} \big| M_i[S_i] \big|^{1/2}} \mathcal{W}_Q(\ud x) 
\end{align*}
where the functionals $\Phi_i(x)$ are
\begin{align*} \nonumber 
& \Phi_i(x):= \frac1{2q_i^2}  \left( \int_{\underline{T}}^{\overline{T}} x[S_i]^{\top} \ud x_i\right) \\ & \hspace{12mm} \cdot \left(M_i[S_i]^{-1}+\frac1{q_i}\mathbb{X}[S_i]\right)^{-1}\left( \int_{\underline{T}}^{\overline{T}} x[S_i] \ud x_i\right),
\end{align*}
$\mathcal{W}_Q(\ud x)$ is the Wiener measure with incremental covariance matrix $Q = \textup{diag}(q_1,...,q_n)$ corresponding to noise process $u_k^{(j)}$ in \eqref{eq:lin_dyn}, and $\mathbb{X}=\int_{\underline{T}}^{\overline{T}} x(t)x(t)^{\top} \ud t$ is the Gramian matrix.
The notation $x[S_i]$ where $S_i$ is the $i^{\textup{th}}$ row of $S$, means the subvector of $x$ in $\mathbb{R}^{|S_i|_0}$ that consists of those elements $x_j$ for which $S_{i,j} = 1$, and for a matrix $K \in \mathbb{R}^{n \times n}$, the notation $K[S_i]$ stands for the $|S_i|_0 \times |S_i|_0$ submatrix of $K$ consisting of those rows and columns of $K$ for which $S_{i,j}=1$.

The integrals with respect to $x$ and $\tau$ are carried out by MCMC sampling. The prior for the pseudotime is a normal distribution. For measurement $k$ done at time $T_j$, we set $p\big(\tau_k^{(j)}|T\big)=N(T_j,\sigma_{\tau})$ (truncated so that $\tau_k^{(j)} \in [\underline{T},\overline{T}]$). Also the covariance parameters $Q = \textup{diag}(q_1,...,q_n)$ and $R = \textup{diag}(r_1,...,r_n)$ are sampled, as well as the indicator matrices $S$, for which the prior is $p(S) \propto \rho^{|S|_0}$ where $\rho \in (0,1)$ is a parameter controlling the sparsity of the samples. For the average of these samples $S^{(j)}$, it holds that
\[
\frac1{L} \sum_{j=1}^L S^{(j)} \to \mathbb{E}(S|Y,T), \qquad \textup{as } L \to \infty
\]
and this average is the output of the algorithm. Since $S$ is a Boolean variable, the elements of $\Ex(S|Y,T)$ are actually the posterior probabilities that the corresponding elements in $A$ are nonzero. The details on the practical implementation of the MCMC sampler as well as details on the computation of the above integrals can be found in \cite{Bayesian_var}.

\subsection{Batch average tracking}

As opposed to novel single-cell techniques, older batch sequencing techniques are only able to provide measurements from population averages. Corresponding to such setup, the method developed in \cite{Bayesian_var} is also included in the comparison, using time series data (with length $m+1$) obtained from the population means
\[
y_j=\frac1{N_j}\sum_{k=1}^{N_j} y_k^{(j)}
\]
with measurement times $T_j$, for $j=0,...,m$.

\section{Numerical experiments} \label{sec:example}

\begin{figure}
\center
\begin{picture}(140,116)
\thicklines
\put(25,25){\circle*{5}}
\put(25,25){\vector(1,0){33}}
\put(25,25){\line(-1,0){15}}
\put(10,25){\line(0,1){35}}
\put(10,60){\line(1,0){11}}
\put(21,56){\line(0,1){8}}
\put(25,25){\line(0,-1){16}}
\put(25,9){\line(1,0){105}}
\put(130,9){\vector(0,1){14}}
\put(25,60){\circle*{5}}
\put(25,60){\vector(0,-1){33}}
\put(25,95){\circle*{5}}
\put(25,95){\vector(0,-1){33}}
\put(60,25){\circle*{5}}
\put(60,25){\vector(1,0){33}}
\put(60,95){\circle*{5}}
\put(60,95){\vector(-1,0){33}}
\put(95,25){\circle*{5}}
\put(95,25){\vector(1,0){33}}
\put(95,95){\circle*{5}}
\put(95,95){\vector(-1,0){33}}
\put(95,95){\line(0,1){15}}
\put(95,110){\line(1,0){50}}
\put(145,110){\line(0,-1){50}}
\put(145,60){\line(-1,0){11}}
\put(134,56){\line(0,1){8}}
\put(130,25){\circle*{5}}
\put(130,25){\vector(0,1){33}}
\put(130,60){\circle*{5}}
\put(130,60){\vector(0,1){33}}
\put(130,95){\circle*{5}}
\put(130,95){\vector(-1,0){33}}
\put(130,60){\line(-1,0){70}}
\put(60,60){\vector(-1,1){33}}
\put(15,95){$1$}
\put(15,65){$2$}
\put(15,13){$3$}
\put(57,13){$4$}
\put(92,13){$5$}
\put(136,22){$6$}
\put(136,65){$7$}
\put(136,95){$8$}
\put(86,100){$9$}
\put(55,100){$10$}
\end{picture}
\caption{The gene regulatory network corresponding to the matrix $A$ of the example. Arrows denote positive effects and blunt arrows denote negative effects.}
\label{fig:GRN}
\end{figure}
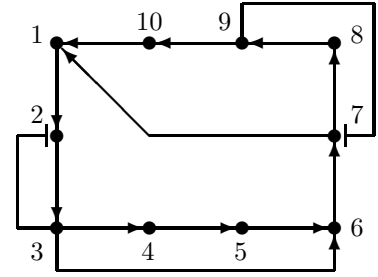

To generate the experimental data, equation \eqref{eq:lin_dyn} was numerically simulated separately for each cell. The used dynamics matrix was
\begin{align*}
&A= \\ &\bm{
 -1 &  0 &  0 &  0 &  0 &  0 &  1 &  0 &  0  &  1 \\
  1 & -1 & -2 &  0 &  0 &  0 &  0 &  0 &  0  &  0 \\
  0 &  1 &  0 &  0 &  0 &  0 &  0 &  0 &  0  &  0 \\
  0 &  0 &  1 & -1 &  0 &  0 &  0 &  0 &  0  &  0 \\
  0 &  0 &  0 &  1 & -1 &  0 &  0 &  0 &  0  &  0 \\
  0 &  0 &  1 &  0 &  1 & -1 &  0 &  0 &  0  &  0 \\
  0 &  0 &  0 &  0 &  0 &  1 & -2 &  0 & -1  &  0 \\
  0 &  0 &  0 &  0 &  0 &  0 &  1 & -1 &  0  &  0 \\
  0 &  0 &  0 &  0 &  0 &  0 &  0 &  1 &  0  &  0 \\
  0 &  0 &  0 &  0 &  0 &  0 &  0 &  0 &  1  & -1}
\end{align*}
corresponding to the gene regulatory network shown in Figure~\ref{fig:GRN}. 
The diagonal elements are chosen so that each column sum is zero. 
The development rates were drawn from a uniform distribution $\gamma_k^{(j)} \sim U(1,1.2)$. The driving Brownian motion $u_k^{(j)}$ had incremental covariance $0.01I$.

The methods were compared in the task corresponding to gene regulatory network inference, that is, inference of the zero structure of the matrix $A$. Well-known classifier scores are used in the comparison, namely the area under the receiver operating characteristic curve (AUROC) and the area under the precision--recall curve (AUPR), excluding the diagonal elements. For the computation of the AUROC and AUPR scores, the methods need to rank the potential links (elements in the $A$ matrix) in the order of confidence. For the distribution-based method, the confidence ranking is obtained simply by ordering the elements of $A$ in decreasing order of their absolute values. This comparison is not entirely fair to the distribution-based method, since it is actually estimating the matrix $A$, rather than the probabilities for the entries being nonzero like the other two methods.

\subsection{Experiment 1}

\begin{table}[b]
\footnotesize
\caption{Measurement times $T_j$ and sizes of measured sub-populations $N_j$ in the different experiments.}
\center
  \begin{tabular}{l l l l l l l l l l}
    \hline 
Exp. 1& $j$ & 0 & 1 & 2 & 3 & 4 & 5 & 6 & 7 \\
Exp. 2--4a & $j$ & 0 &  & 1 & 2 & & 3 & & 4 \\
 \hline
& $T_j$ & 0 & 0.2 & 0.5 & 1.2 & 2.2 & 2.95 & 4 & 5.2 \\
& $N_j$ & 50 & 59 & 57 & 55 & 54 & 64 & 60 & 46 \\
    \hline
  \end{tabular}
  \label{tab:data}
\end{table}


\begin{figure*}[t]
\center
\mbox{
\hspace{-3mm}
\includegraphics[width=6.2cm]{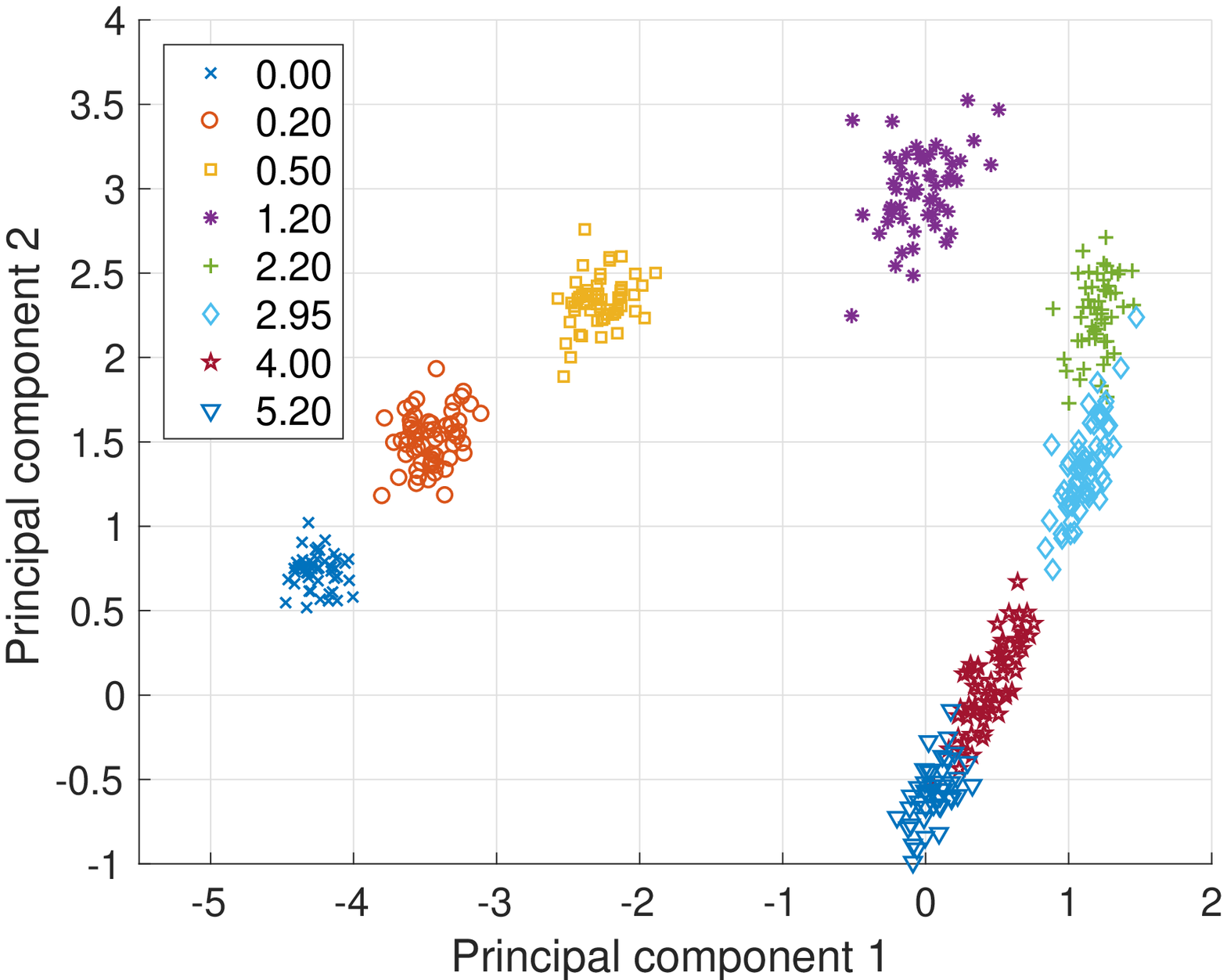}
\hspace{-5mm}
\includegraphics[width=6.2cm]{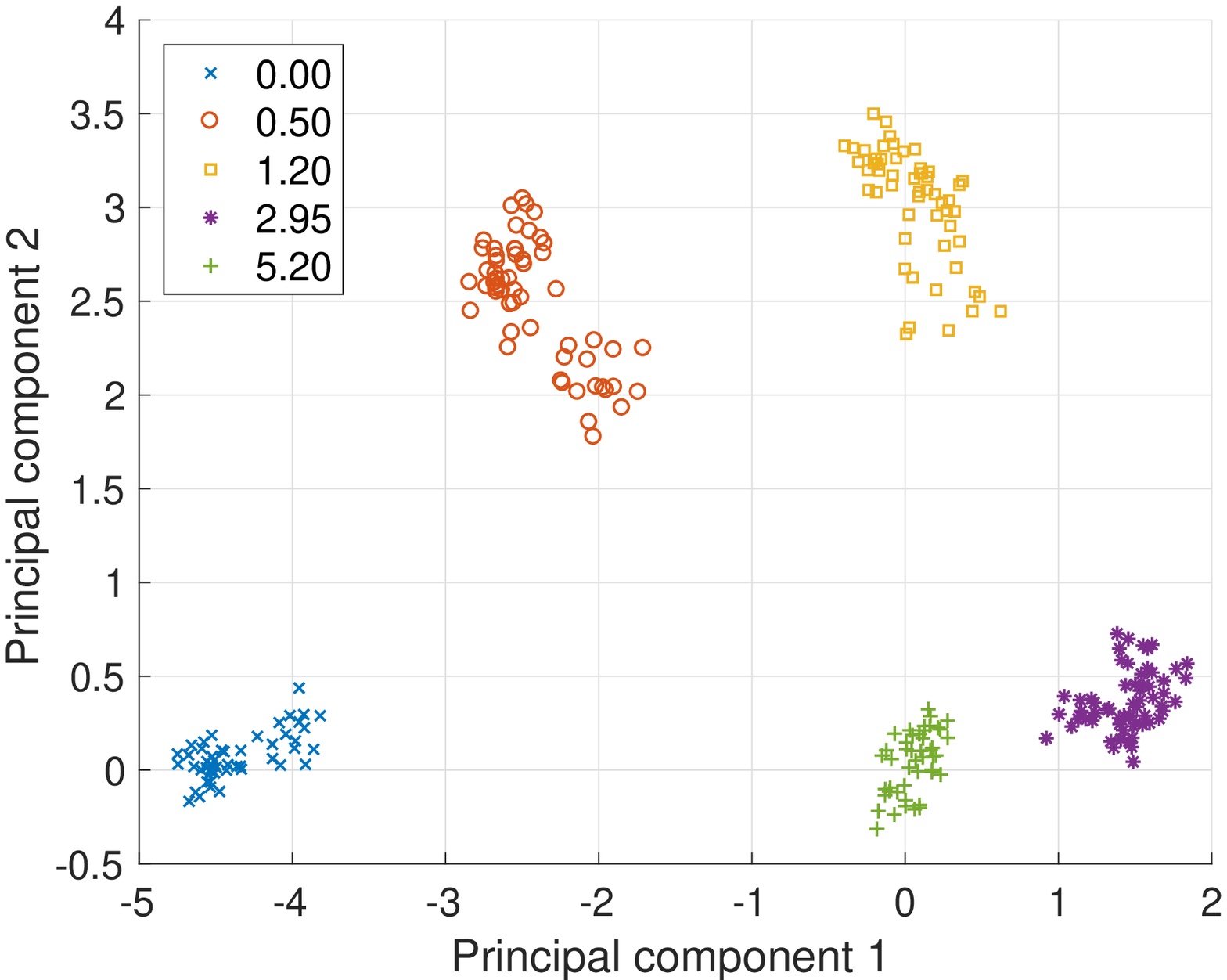}
\hspace{-5mm}
\includegraphics[width=6.2cm]{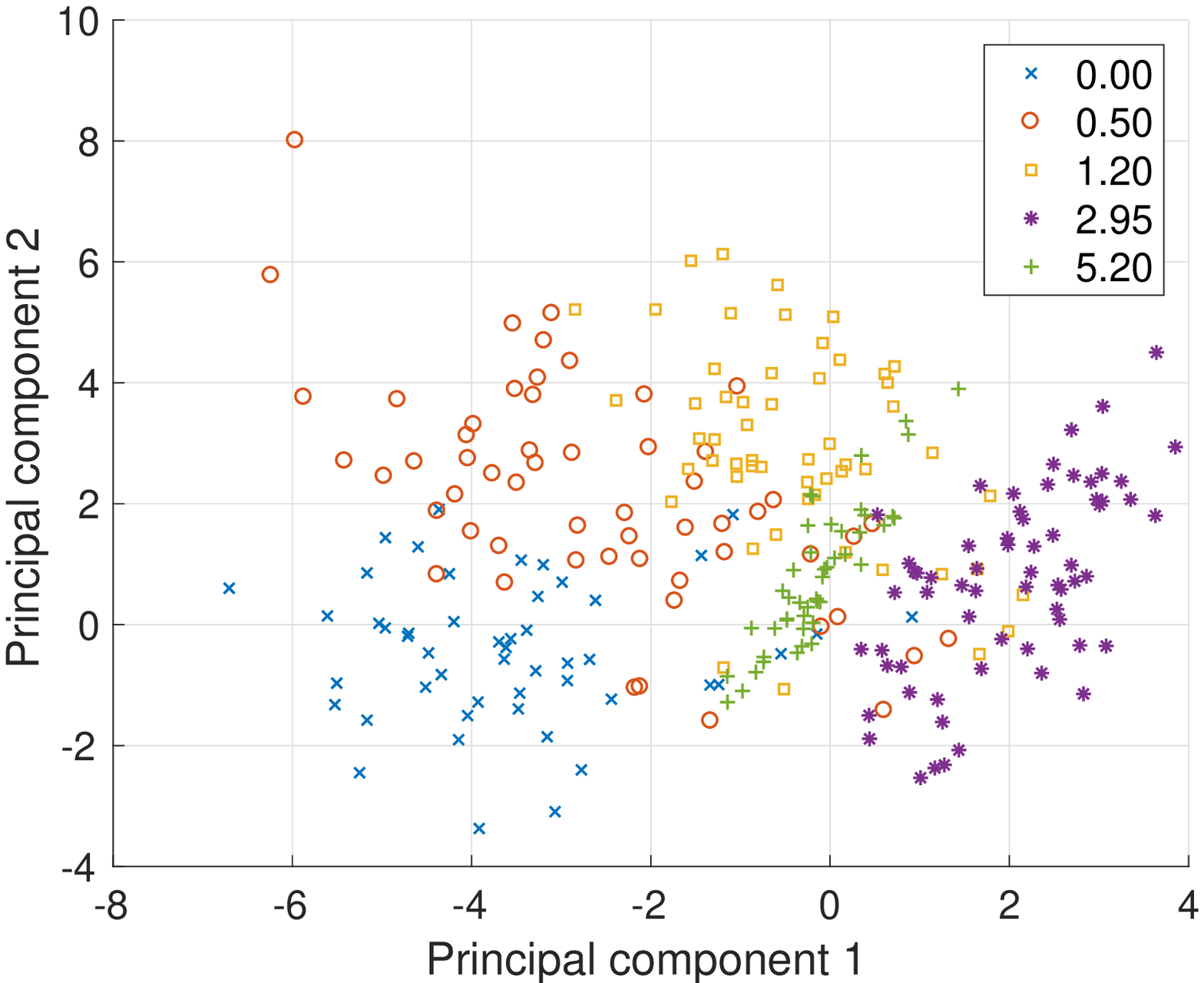} }
\caption{The first two principal components of the simulated data for experiments 1 and~2 (left), experiment 3 (center), and experiment 4a (right). Each point corresponds to one measured cell. Different measurement times are indicated with different symbols and colours.}
\label{fig:data}
\end{figure*}

In the first experiment, altogether 445 measurements are collected at eight different times, as described in Table~\ref{tab:data}. The initial distribution is a normal distribution, $P_0 = N(m_0,\Gamma)$ where $m_0$ was also randomly chosen and $\Gamma = \textup{diag}(.1,.05,.16,.2,.11,.19,.18,.07,.11,.09)^2$. The data are visualised in Figure~\ref{fig:data} (left) showing first two principal components. From the figure it can be seen that the later measurements are more spread in the direction of the main propagation due to the different development rates.

The distribution-based method was tested with six different values of $\lambda$, which is the cost function parameter penalising for the 1-norm of the $A$-matrix. Similarly, the two other methods were tried with six different values of the sparsity parameter $\rho$. The resulting AUROC and AUPR values (for four interesting parameter values) are shown in Table~\ref{tab:results}. The pseudotime method shows a more solid performance than the distribution-based method, even obtaining perfect reconstruction with $\rho=0.3$. The batch-method {\it C} is clearly the weakest, which is not surprising.

\begin{figure}[b]
\center
\includegraphics[width=8cm]{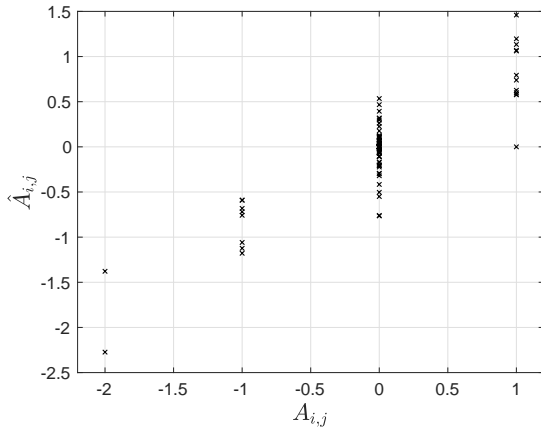} 
\caption{The elements of $A$ estimated with the distribution-based method in experiment 3 plotted against the true values.}
\label{fig:Ahat}\end{figure}

\subsection{Experiment 2}

In the second experiment, the amount of data was reduced by using only five of the eight populations in Experiment~1 (see Table~\ref{tab:data} and Figure~\ref{fig:data} (left)) resulting in 272 measurements. In this experiment, only the values $\lambda=0.005$ and $\rho=0.2$ were used. The results are shown in Table~\ref{tab:results}.

Again, the pseudotime method {\it B} was better than the distribution-based method {\it A}. However, the pseudotime method seemed to be sensitive to the initialisation of the continuous trajectory $x$ in the sampler, and sometimes the MCMC sampler converged to the neighbourhood of a local maximum of the posterior distribution, which did not yield as good results as those reported in Table~\ref{tab:results}. Since the batch-method {\it C} only has five data points in this experiment, its performance is clearly worse than in Experiment~1.

\subsection{Experiment 3}


In the third experiment, the initial distribution $P_0$ was a mixture of two Gaussians, so that with probability 0.7, the initial state $x_k^{(j)}(0)$ was drawn from normal distribution $N(m_0,\Gamma)$, and with probability 0.3, the initial state was drawn from $N(m_1,\Gamma)$, where $m_0$ and $m_1$ were close to each other. This experiment is simulating heterogeneity in the cell population. The measurement times and population sizes are the same as in Experiment~2. The data are visualised in Figure~\ref{fig:data} (center). The resulting AUROC/AUPR values are in Table~\ref{tab:results}, and the entries of matrix $A$ estimated with the distribution-based method are visualised in Figure~\ref{fig:Ahat}. This time the distribution-based method outperforms the pseudotime-based method, and it even attains higher AUROC/AUPR scores than in Experiment~2. This result is expected, since the distribution-based method gets more information from the heterogeneity in the distribution, whereas the pseudotime method erroneously tries to fit the heterogeneity by adjusting the pseudotimes.

\subsection{Experiment 4}

In the fourth experiment the cell variability was more realistic. The initial distribution was again a mixture of two Gaussian distributions, with the distance between their means doubled compared to experiment 3, and their covariances were 100$\Gamma$. 
 In experiment 4a, the amount of data is as in Table~\ref{tab:data}, and these data are visualised in Figure~\ref{fig:data} (right). In experiment 4b, the amount of measured cells at each time point was doubled, resulting in 544 measurements in total, at five different times.

With smaller amount of data in experiment 4a, the best results were surprisingly obtained by method {\it C}, implying that the other methods were unable to obtain meaningful information from the measurement distributions. Method {\it B} suffered again of multimodality problems in MCMC sampling and the results were gathered from five independent sampling chains. On the other hand, when the number of measurements was increased in experiment 4b, then the distribution-based method was again the best performer.


It should be noted that with linear systems, the mean of population $j$ is propagated by $e^{A(T_{j+1}-T_j)}$ to the mean of population $j+1$. This is not true with nonlinear systems, and therefore a method tracking the averages of the measured batches is likely to perform worse with nonlinear systems.

\begin{table*}[t]
\footnotesize
\vspace{1mm}
\caption{AUROC/AUPR values for the methods {\it A}: the distribution-based method, {\it B}: the pseudotime method, and {\it C}: the batch average method. Each method has some sparsity-enforcing parameter, and the results were established with different values of these parameters. Note that higher $\lambda$ promotes sparser solutions, whereas higher $\rho$ promotes less sparse solutions.}
\center
  \begin{tabular}{|l l | r@{\hspace{.4mm}/\hspace{.4mm}}l r@{\hspace{.4mm}/\hspace{.4mm}}l r@{\hspace{.4mm}/\hspace{.4mm}}l r@{\hspace{.4mm}/\hspace{.4mm}}l | r@{\hspace{.4mm}/\hspace{.4mm}}l | r@{\hspace{.4mm}/\hspace{.4mm}}l | r@{\hspace{.4mm}/\hspace{.4mm}}l  r@{\hspace{.4mm}/\hspace{.4mm}}l |}
    \hline 
    Experiment & & \multicolumn{8}{|c}{1} & \multicolumn{2}{|c|}{2} & \multicolumn{2}{|c|}{3} & \multicolumn{2}{|c}{4a} & \multicolumn{2}{c|}{4b} \\ \hline
  Parameter & $\lambda$  & \multicolumn{2}{c}{0.0025} & \multicolumn{2}{c}{0.005} & \multicolumn{2}{c}{0.01} & \multicolumn{2}{c}{0.05} & \multicolumn{2}{|c|}{0.005} & \multicolumn{2}{c|}{0.005} & \multicolumn{2}{c}{0.005} & \multicolumn{2}{c|}{0.005}  \\ 
Method & {\it A} & {0.923} & 0.881 & 0.931 & 0.900 & {\bf 0.984} & {\bf 0.941} & 0.931 & 0.888 & 0.751 & 0.589 & 0.927 & 0.868 & 0.661 & 0.332 & 0.913 & 0.837 \\ \hline
 Parameter & $\rho$ &  \multicolumn{2}{c}{0.3} &  \multicolumn{2}{c}{0.25} & \multicolumn{2}{c}{0.2} & \multicolumn{2}{c}{0.15}  & \multicolumn{2}{|c}{0.2} & \multicolumn{2}{|c}{0.2} & 
 \multicolumn{2}{|c}{0.2} & \multicolumn{2}{c|}{0.2} \\
Method & {\it B} & {\bf 1.000} & {\bf 1.000} & 0.977 & 0.899 & 0.965 & 0.879 & {0.981} & {0.909} &  0.862 & 0.639 & 0.845 & 0.622 & 0.733 & 0.389 & 0.895 & 0.753 \\ 
Method & {\it C} & 0.852 & 0.696 & {\bf 0.864} & {\bf 0.716} & 0.856 & 0.712 & 0.842 & 0.696 & 0.723 & 0.310 & 0.762 & 0.469 & 0.761 & 0.470 & 0.745 & 0.404  \\
    \hline
  \end{tabular}
  \label{tab:results}
\end{table*}

\section{Conclusions}
Two different paradigms were introduced for identifying linear systems from snapshot ensemble observation data. The first paradigm is based on tracking the evolution of the distributions of cells accross time. The second paradigm is based on the pseudotime concept, where the idea is based on the fact that the cells evolve with different rates and therefore one snapshot contains information from different development stages of the cell.
The developed pseudotime-method samples trajectories from which the measurements are obtained at different (pseudo)times (since the cells develop with different rates). On average, the pseudotime-method gave slightly better results than the distribution-based method, and when the data contained only moderate noise (and no model class mismatch), its performance was excellent. However, the distribution-based method seemed to be more robust against disturbances. The pseudotime-method tries to fit the trajectory and the pseudotimes into the data. If the data contain some systemic heterogeneity which is not due to the varying developmental rates (such as in Experiment~3), then the method will try to explain the heterogeneity by the pseudotimes, causing an error in the method. Some pseudotime estimation methods are able do detect branches in the biological processes \cite{Branch_GP}. In such approach, one trajectory only takes into account data belonging to one branch, thus avoiding overfitting. Obviously, the relative performances of the methods may still vary depending on the quality of the data. One observation is that when the cell variability is high (Experiment~4), then sufficiently many measurements are needed in order to obtain information from the distribution of cells. On the other hand, single-cell experimental techniques are developing fast, and the number of measured cells in most experiments far exceeds what we used in the numerical experiments.

Future work includes implementation of nonlinear dynamics either using a mechanistic approach \cite{Mechanistic_SC_GRN} or nonparametric dynamics functions \cite{GPDM_GRN}, development of a more efficient optimisation scheme for solving \eqref{eq:opt_prob}, and experiments using real data.

%
%
%



\def\url#1{}      
\bibliographystyle{IEEEtran}

\end{document}